\DocumentMetadata{}
\documentclass[sigplan,screen]{acmart}

\AtBeginDocument{%
  \providecommand\BibTeX{{%
    \normalfont B\kern-0.5em{\scshape i\kern-0.25em b}\kern-0.8em\TeX}}}

\begin{document}

\title{Structure-Aware Code Vulnerability Analysis With Graph Neural Networks}

\author{Ravil Mussabayev}
\email{ravmus@gmail.com}
\affiliation{%
  \institution{Moscow Software Development Tools Could Technology Lab,\\Huawei Russian Research Institute}
  \streetaddress{Smolenskaya Square 5}
  \city{Moscow}
  \country{Russia}
  \postcode{121099}
}

\begin{abstract}
  This study explores the effectiveness of graph neural networks (GNNs) for vulnerability detection in software code, utilizing a real-world dataset of Java vulnerability-fixing commits. The dataset's structure, based on the number of modified methods in each commit, offers a natural partition that facilitates diverse investigative scenarios. The primary focus is to evaluate the general applicability of GNNs in identifying vulnerable code segments and distinguishing these from their fixed versions, as well as from random non-vulnerable code. Through a series of experiments, the research addresses key questions about the suitability of different configurations and subsets of data in enhancing the prediction accuracy of GNN models. Experiments indicate that certain model configurations, such as the pruning of specific graph elements and the exclusion of certain types of code representation, significantly improve performance. Additionally, the study highlights the importance of including random data in training to optimize the detection capabilities of GNNs.
\end{abstract}


\maketitle

\section{Introduction}

Code vulnerability detection is a critical challenge in software security that has significant implications for both individuals and organizations \cite{Fu2022}. As software systems grow increasingly complex and interconnected, the presence of vulnerabilities poses serious threats, including potential breaches, data leaks, and compromised user privacy. Detecting code vulnerabilities is essential to proactively identify and remediate security flaws before they can be exploited by malicious actors. However, manual inspection of code for vulnerabilities is time-consuming, error-prone, and impractical for large-scale codebases. Therefore, developing automated methods, such as machine learning models, for accurately and efficiently identifying code vulnerabilities is of paramount importance to enhance software security and protect against potential risks.

\begin{figure*}
\centering
\includegraphics[scale=0.4]{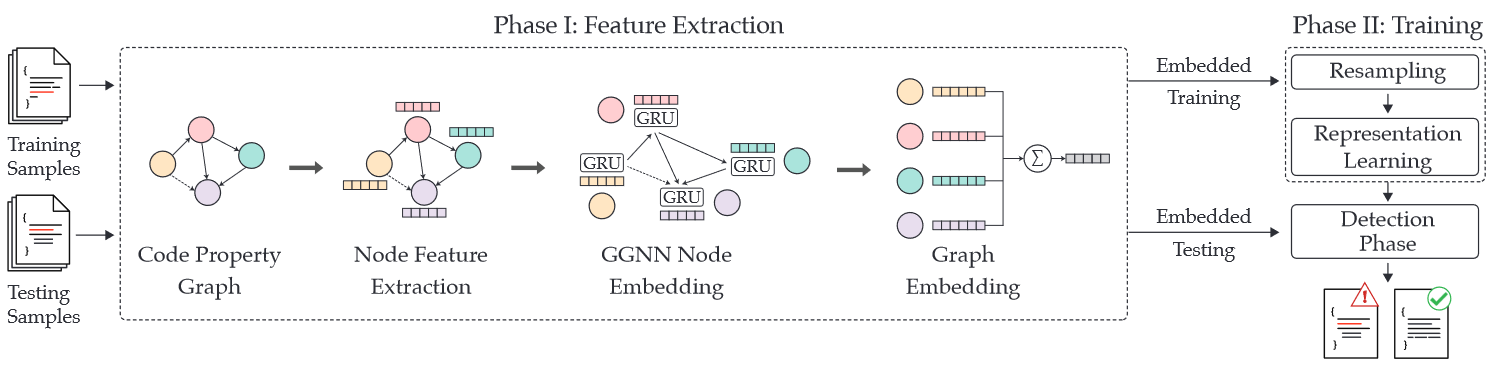}
\caption{Architecture of the ReVeal model}
\label{fig:reveal_arch}
\end{figure*}

One of the state-of-the-art models for vulnerability detection is ReVeal \cite{Chakraborty2022}. It consists of three main modules: a gated graph neural network (GGNN), a SMOTE resampling, and a representation learning block. For a pictorial representation of the architecture, see Figure \ref{fig:reveal_arch}. The authors collected a new dataset of C++ vulnerabilities from the Linux Debian Kernel and the Chromium projects. In each of the security patches, they annotated the previous versions of all changed functions (i.e., the versions prior to the patch) as ``vulnerable'' and the fixed version of all changed functions (i.e., the version after the patch) as ``clean''. Additionally, other functions that were not involved in the patch (i.e., those that remained unchanged) are all annotated as ``clean''. From now on, this dataset will be referred to as the ``ReVeal'' dataset.

In \cite{Chakraborty2022}, extensive experiments with other vulnerability detection models present in the literature showed their acute inadequacy when tested on the ReVeal dataset. ReVeal dataset stands out from other vulnerability detection benchmarks available in the literature. It consists of real imbalanced data annotated by human developers. Other datasets were mainly synthetic or semi-synthetic annotated by static analysers or unsupervised techniques. A misleadingly high scores achieved by other models could be explained by the low quality of their test datasets. More concretely, the authors of \cite{Chakraborty2022} point at the following limitations of other existing approaches: (a) data duplication, (b) not handling data imbalance, (c) not learning semantic information, (d) lacking class separability.

In this study, we would like to reproduce the results of \cite{Chakraborty2022} and answer the following research question:

\textbf{Research Question 1}. How would one optimize the parameters in the following dimensions to achieve the best possible performance of the ReVeal model on the ReVeal dataset?
\begin{enumerate}
\item Include the SMOTE and representation learning modules or only use a single GGNN block;
\item Include information about AST edges into the input graphs or only use DDG and CFG edges;
\item Include the full graph, which can be too large and detailed, or use its pruned version at operator nodes instead;
\item Balance the training data by downsampling the majority class or keep the original class ratio.
\end{enumerate}

We also investigate the performance of the ReVeal model on the Java vulnerability detection data. We collected a large dataset of $865$ Java vulnerability-fixing commits across a wide variety of CWE vulnerability types. In this work, we restrict ourselves to the problem of method-level vulnerability detection. Each commit can involve a different number of changed methods. This induces a natural partition of the dataset with respect to the number of changed functions in a commit.

Statistically, there are many more commits where more than one function has changed. Thus, a crucial problem arises in this setting. In a vulnerability-fixing commit where more than one function has changed, we cannot ensure that all the changes are related to fixing the underlying vulnerability. Thus, not all matched pairs of functions involved in such a commit can be labelled as fixing the vulnerability.

To address the above-mentioned issue, we compile and answer a list of research questions. To formulate these questions, we introduce a new notation. We create a sequence of sets with the following structure:
\vspace{10pt}
\begin{align}
  D_k = & \ P_1 \cup P_2 \cup P_3 \label{eq:d_k} \\ \nonumber
      = & \ \{ (f, f') \in C \ | \ C \ \text{is strict} \} \ \cup \\ \nonumber
        & \ \{ (f, f') \in C \ | \ C \ \text{is $k$-strict} \} \ \cup \\ \nonumber
        & \ \{ f \ | \ f \ \text{is random \& safe} \} \nonumber
\end{align}
where $C$ is a vulnerability-fixing commit, and $(f, f')$ denotes the pair of an original function $f$ with its changed version $f'$. We say that a vulnerability fixing commit $C$ is $k$-strict if it contains exactly $k$ changed pairs of functions.

We also decompose the original problem into two independent tasks:
\begin{itemize}
\item Task $T_1$: dividing in the set of potentially vulnerable methods, i.e., vulnerable methods (active vulnerable) against their fixed versions (passive vulnerable);
\item Task $T_2$: dividing in the set of all methods, i.e., potentially vulnerable methods against random safe code.
\end{itemize}
Then, the zero-day vulnerability detection task is a composition of these two tasks
$$
T_0 = T_1 \circ T_2
$$

Our hypothesis is that task $T_1$ is much more difficult than task $T_2$ for the state-of-the-art models. If we answer this question in the positive, then task $T_1$ is a bottleneck and there is a need to develop better models that are specifically tailored to tackle $T_1$.

Thus, we have the following list of research questions:

\textbf{Research Question 2}. Is $P_i$ useful? ($i = 1, 2, 3$)

\textbf{Research Question 3}. How difficult is task $T_1$?

\textbf{Research Question 4}. How difficult is task $T_2$?

\textbf{Research Question 5}. Does random code appear in $P_2$ as $k$ increases?

\textbf{Research Question 6}. How does the size of $P_3$ affect the overall performance?

\section{Experimental setup}

We adapted the source code from GitHub provided by the authors of \cite{Chakraborty2022}: \href{https://github.com/VulDetProject/ReVeal}{https://github.com/VulDetProject/ReVeal}. The experiments were conducted on a computer with the following configuration: Intel(R) Xeon(R) Gold 6151 CPU @ 3.00GHz with 32 cores, NVIDIA Tesla V100 PCIe GPU with 16 GB, and 126 GB RAM. The computing platform had the following specifications: Python 3.10.7, NumPy 1.23.3, DGL 1.0.1+cu117. Joern of version 1.1.1495 was used to parse source code into a graph representation.

Throughout the experiments, we used the default choice of hyperparameters: learning rate $0.0001$, weight decay $0.001$, graph embedding size $200$, batch size $128$, maximum number of batches $10000$, number of gradient accumulation steps $8$, maximum patience of $50$ for C++ data and $20$ for Java data.

\section{Experiments with C++ data}

To answer the first research question, we used the original C++ method-level vulnerability dataset from \cite{Chakraborty2022}. After parsing, we obtained the following statistics of the input graphs:

\begin{center}
11788 train graphs (956 vulnerable), 1667 validation graphs (133 vulnerable), 3385 test graphs (286 vulnerable)
\end{center}

To test each dimension of RQ 1, we performed 10 trials of training the model. In each trial, the dataset was split into train, validation, and test parts anew. The results can be found in Table \ref{tab:rq1}.

\begin{table}[!htbp]
    \centering
    \begin{tabular}{c|cc}
         Configuration & Median F1 & Median ROC AUC \\ \hline
         Baseline & 27.29 & 0.696 \\
         Without SMOTE \& RL & 21.45 & 0.730 \\
         Without AST edges & 27.65 & 0.706 \\
         With pruning & 30.83 & 0.724 \\
         Majority downsampling & 26.61 & 0.678
    \end{tabular}
    \caption{Results of experiments for research question 1}
    \label{tab:rq1}
\end{table}

\subsection{Excluding SMOTE and RL}

The model without SMOTE and RL achieves the worst performance with respect to the F1 score and the best performance with respect the ROC AUC measure.

\subsection{AST edges}

The model performs slightly better without including AST edges. This is likely due to including too much of fine-grained information or too many nodes. The model becomes more likely to overfit to irrelevant features in the input and fail to generalize.

\subsection{Pruning}

The experiments also showed that the model performs better with pruning at operator nodes. Pruning makes a graph simpler and less entangled for the model to understand.

\subsection{Downsampling}

Table \ref{tab:rq1} shows that the model performs worse with balancing the train set by downsampling non-vulnerable methods. We think that a rough balancing of the train part impacts the score negatively since it turns off SMOTE.

\section{Experiments with Java data}

To answer the rest of research questions, we trained and tested the model on different parts of the Java dataset \eqref{eq:d_k}: $P_1$, $P_2$, and $P_3$. In particular, we varied $k$ in the range from $1$ to $14$. Then, we plotted the resulting ROC AUC scores against $k$, and draw conclusions based on the observed dynamics. To make set $P_3$ to be independent of $k$, we fixed it to be the complement of $P_1$. That is, $P_3$ consisted of functions that remained unchanged in the commits where only one function was changed. Also, in order to balance different parts involved in training and testing, we restricted the size of $P_3$:
$$
|P_3| = |P_1| + |P_2|
$$

During the data cleaning phase, we ensured that in each experiment, $P_3$ did not contain functions that are contained in $P_1 \cup P_2$. Also, we removed any duplicate functions from each of the parts $P_1, P_2$, and $P_3$, and removed methods contained in the training data from the test data.

Table \ref{tab:java_stat} shows the distribution of the collected Java methods after stratification by $k$ and cleaning the data:

\begin{table}[!htbp]
    \centering
        \begin{tabular}{|c|c|c|c|c|}
        \hline
         & \multicolumn{2}{c|}{P1} & \multicolumn{2}{c|}{P2} \\
        \hline
        k & train & test & train & test \\
        \hline
        1 & 410 (205) & 135 (68) & 0 (0) & 0 (0) \\
        2 & 399 (200) & 145 (73) & 343 (171) & 122 (61) \\
        3 & 416 (208) & 132 (66) & 696 (347) & 228 (113) \\
        4 & 414 (207) & 128 (65) & 960 (479) & 346 (172) \\
        5 & 415 (210) & 129 (64) & 1159 (575) & 433 (217) \\
        6 & 414 (208) & 131 (65) & 1393 (692) & 506 (254) \\
        7 & 421 (212) & 120 (60) & 1583 (789) & 596 (296) \\
        8 & 394 (197) & 151 (75) & 1870 (938) & 572 (284) \\
        9 & 410 (207) & 135 (67) & 2027 (1012) & 664 (330) \\
        10 & 411 (206) & 131 (66) & 2195 (1089) & 632 (314) \\
        11 & 399 (199) & 150 (75) & 2439 (1215) & 708 (353) \\
        12 & 400 (202) & 144 (72) & 2545 (1270) & 769 (383) \\
        13 & 397 (200) & 143 (72) & 2619 (1303) & 872 (434) \\
        14 & 409 (204) & 136 (68) & 2853 (1421) & 845 (419) \\
        \hline
        \end{tabular}
    \caption{Statistics of collected Java methods after stratification by $k$ and cleaning. Each cell has the format $N_1 (N_2)$, where $N_1$ is the total number of methods and $N_2$ is the number of vulnerable ones.}
    \label{tab:java_stat}
\end{table}

\subsection{Research question 2}

\begin{figure}
    \centering
    \includegraphics[scale=0.6]{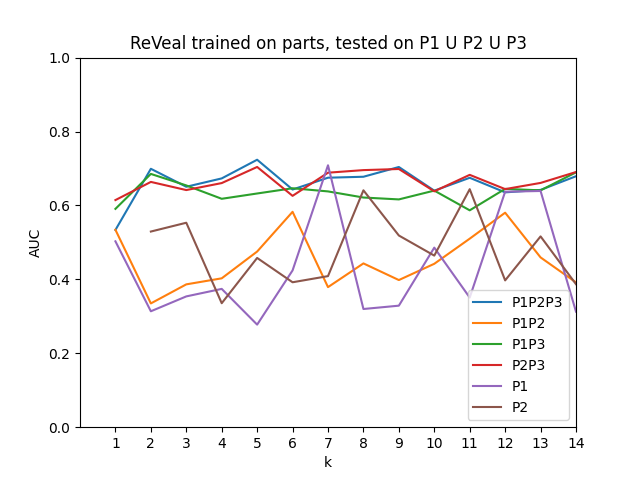}
    \caption{ReVeal model trained on parts, tested on $P_1 \cup P_2 \cup P_3$.}
    \label{fig:aucs_p1p2p3}
\end{figure}

\begin{figure}
    \centering
    \includegraphics[scale=0.6]{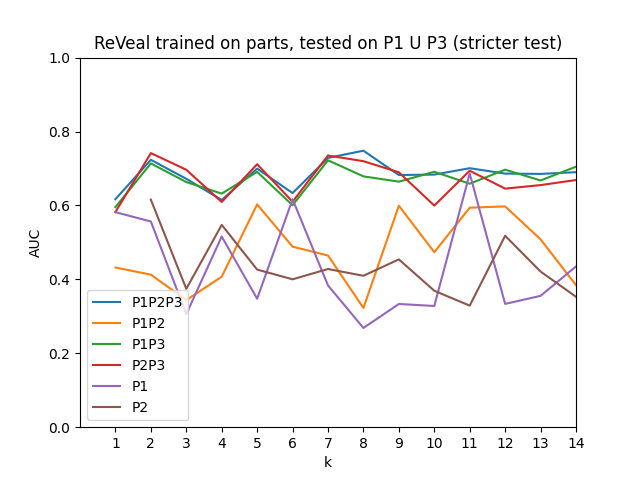}
    \caption{ReVeal model trained on parts, tested on $P_1 \cup P_3$. This is a stricter test for task $T_2$ than the one on Figure \ref{fig:aucs_p1p2p3}).}
    \label{fig:aucs_p1p3}
\end{figure}

In this research question, we investigate training on different combinations of sets $P_1$, $P_2$, and $P_3$, and testing on $P_1 \cup P_2 \cup P_3$ or $P_1 \cup P_3$, which is a stricter test. The results can be found in Figures \ref{fig:aucs_p1p2p3} and \ref{fig:aucs_p1p3}.

Figures \ref{fig:aucs_p1p2p3} and \ref{fig:aucs_p1p3} allow us to conclude that if the test set includes part $P_3$, then the inclusion of part $P_3$ into training is critical to achieving a high performance. Overall, parts $P_2$ and $P_3$ contribute the most to the prediction, as seen by the red and blue lines on Figures \ref{fig:aucs_p1p2p3} and \ref{fig:aucs_p1p3}.

Also, on Figure \ref{fig:aucs_p1p3}, we see a slight degradation of performance corresponding to training on $P_2 \cup P_3$ (red line) as $k$ increases. This might indicate the increasing amount of random noise in $P_2$ as $k$ increases, partially answering RQ 5 in the positive.

\subsection{Research question 3}

\begin{figure}
    \centering
    \includegraphics[scale=0.6]{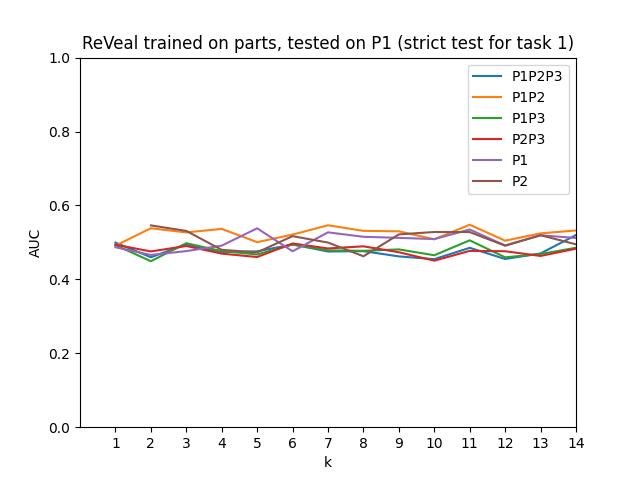}
    \caption{ReVeal model trained on parts, tested on $P_1$. This is a strict test for task $T_1$.}
    \label{fig:aucs_p1}
\end{figure}

To assess the quality of the model on task $T_1$, we change the combination of the test set to $P_1$. This is the strictest possible test set that reflects the ability of the model to distinguish small differences between very similar code. The resulting plot can be found in Figure \ref{fig:aucs_p1}.

As you can infer from this figure, the ReVeal model is unable to perform much better than random guessing on this test data irrespective of the training configuration. However, a slightly better result is achieved by the training data consisting of the set $P_1 \cup P_2$. Also, training on sets $P_1$ and $P_2$ separately shows a performance better than random strategy for most values of $k$. Including data from $P_3$ into training misleads the model since the test data does not have instances from the distribution of $P_3$.

This experiment shows that the ReVeal model heavily underperforms on task $T_1$.

\subsection{Research question 4}

\begin{figure}
    \centering
    \includegraphics[scale=0.6]{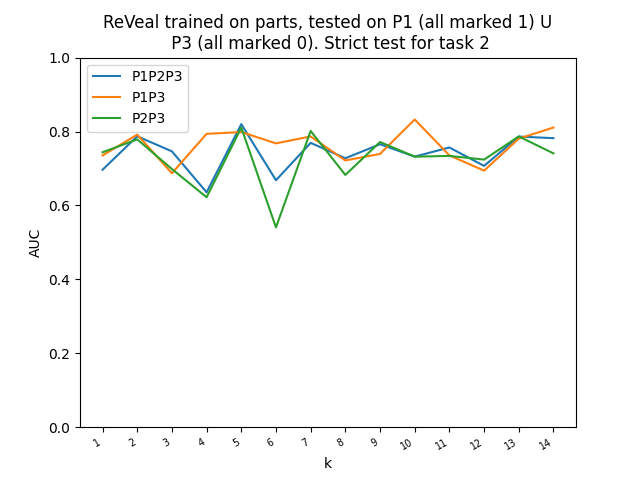}
    \caption{ReVeal model trained on parts, tested on $P_1$ (all marked positive) $\cup$ $P_3$ (all marked negative). This is a strict test for task $T_2$.}
    \label{fig:aucs_te_p1p_p3n}
\end{figure}

In this experiment, the training data consisted of instances from set $P_1 \cup P_2$ marked as a positive class, and instances from part $P_3$ marked as a negative class. Likewise, the test data was comprised of instances from part $P_1$ marked as a positive class, and instances from part $P_3$ marked as a negative class. The results of this setting reflect the ability of the model to differentiate close-to-vulnerable code from random safe code. The resulting ROC AUC values for this experiment can be found in Figure \ref{fig:aucs_te_p1p_p3n}.

From Figure \ref{fig:aucs_te_p1p_p3n}, we see that the model performs fairly well on task $T_2$, achieving the best results in the training regime involving $P_1 \cup P_3$.

\subsection{Research question 5}

\begin{figure}
    \centering
    \includegraphics[scale=0.6]{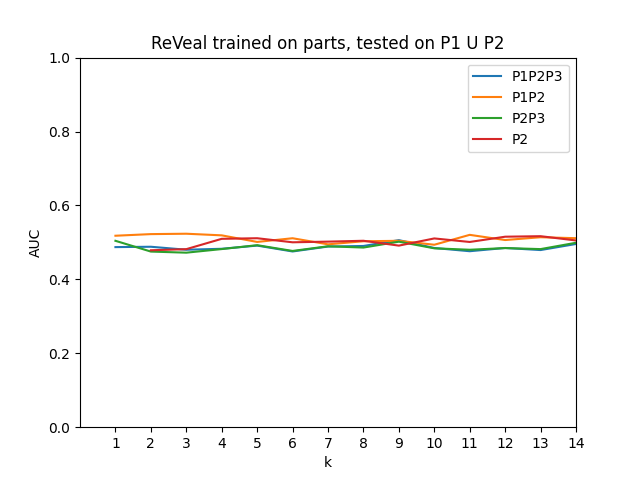}
    \caption{ReVeal model trained on parts, tested on $P_1 \cup P_2$.}
    \label{fig:aucs_te_p1p2}
\end{figure}

To answer this research question, we train on all possible combinations of parts involving $P_2$. Then, we test the trained models on $P_1 \cup P_2$. The results can be seen in Figure \ref{fig:aucs_te_p1p2}.

This plot does not allow us to definitively conclude anything about RQ 5. The model does not perform adequately for any training regime, and does not exhibit any trends that can be detected across different values of $k$.

\subsection{Research question 6}

\begin{figure}
    \centering
    \includegraphics[scale=0.6]{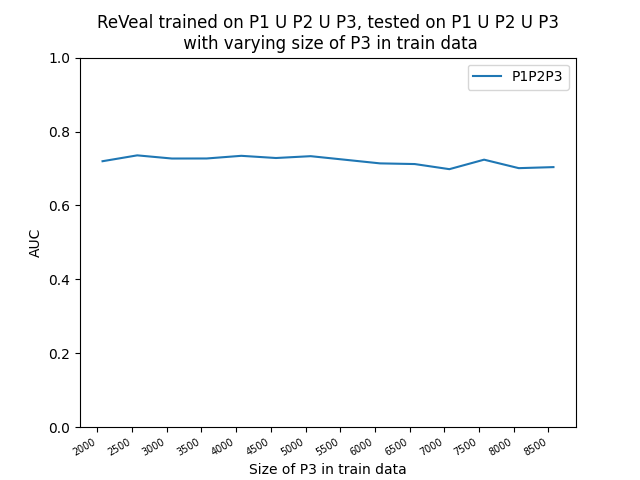}
    \caption{ReVeal model trained on $P_1 \cup P_2 \cup P_3$, tested on $P_1 \cup P_2 \cup P_3$ with varying size of $P_3$ in the train data.}
    \label{fig:aucs_te_p1p2p3_diff_p3_size}
\end{figure}

RQ 2 concluded that including $P_3$ into the training data is crucial for achieving good performance on task $T$. Now, we investigate how the size of $P_3$ in the training data affects the final performance. For that, we pick the best training configuration from RQ 2 and plot its performance against different sizes of $P_3$ in the training data. The test set was fixed. The results are displayed on Figure \ref{fig:aucs_te_p1p2p3_diff_p3_size}.

It is clear from Figure \ref{fig:aucs_te_p1p2p3_diff_p3_size} that the size of $P_3$ in the training data does not affect the performance of the model.

\section{Threats to validity}

There are some threats to the validity of the current study:

\begin{enumerate}
    \item Insufficient number of trials made for each choice of $k$ (for RQs 2--5) and $|P_3|$ (for RQ 6). Only one trial was performed in our study. This might not be enough to account for various random phenomena present in the training of a graph neural network and in the split of data into train, validation, and test subsets;
    \item The available training data might be insufficient to make solid conclusions. In particular, we only had around $3148$ training examples for $k = 5$ in the regime $P_1 \cup P_2 \cup P_3$. This number might not suffice to train a large graph neural network.
\end{enumerate}

\section{Conclusion}

In this study, we have investigated the performance of the ReVeal model on different datasets and configurations, focusing on the identification of vulnerable code in C++ and Java. We used a graph-based approach, representing code as graphs, and employed a machine learning model to predict vulnerabilities.

Our investigation focused on six research questions (RQs), each exploring different aspects of the problem. RQ 1 showed that the model performs better with pruning and without including AST edges, while the absence of SMOTE, RL, and downsampling techniques led to a decrease in performance.

The analysis of RQ 2 and RQ 3 indicated that the inclusion of part $P_3$ in the training set is critical for achieving high performance, especially when the test set also includes part $P_3$. However, the model underperformed on task $T_1$, suggesting that distinguishing small differences between very similar code remains challenging.

Our findings from RQ 4 and RQ 5 provided insights on the model's outstanding performance in differentiating close-to-vulnerable code from random safe code and the possible effect of random noise in $P_2$ as $k$ increases. In RQ 6, we found that the size of $P_3$ in the training data did not significantly affect the model's performance.

While our research offers substantial insights, there are a few threats to its validity. These include the limited number of trials for each choice of $k$ and $|P_3|$, and the possibility that the available training data may be insufficient to make solid conclusions.

In conclusion, our research provides valuable insights into the performance and suitability of graph neural networks for identifying vulnerable code. However, more research is required to further refine these models and address the identified challenges, particularly in distinguishing small differences in code and handling imbalanced datasets. Future work could involve investigating other machine learning models, as well as improving data augmentation techniques to enhance the model's performance.

\bibliographystyle{ACM-Reference-Format}
\bibliography{main}

\appendix

\end{document}